\documentclass[twocolumn,showpacs,amsmath,amstex,amssymb,mathfonts,prl,superscriptaddress]{revtex4}
\usepackage{amsthm,amsfonts,graphicx,verbatim}
\usepackage{amsmath}
\usepackage{amssymb}
\usepackage{amsthm}
\usepackage{amsfonts}
\usepackage{listings}
\usepackage{enumerate}
\usepackage{latexsym}
\usepackage{psfrag}
\usepackage{bm}
\usepackage{graphicx}
\usepackage{subfigure}
\newcommand{\be}{\begin{equation}}
\newcommand{\ee}{\end{equation}}
\newcommand{\bea}{\begin{eqnarray}}
\newcommand{\eea}{\end{eqnarray}}

\newcommand{\la}{\langle}
\newcommand{\ra}{\rangle}

\renewcommand{\phi}{\varphi}
\renewcommand{\epsilon}{\varepsilon}

\begin{document}

\title{Interaction-driven topological insulator states in strained graphene}

\author{D. A. Abanin}
\affiliation{Department of Physics, Harvard University, Cambridge, Massachusetts 02138, USA}
\author{D. A. Pesin}
\affiliation{Department of Physics, University of Texas at Austin,  Austin TX 78712, USA}


\date{\today}

\begin{abstract}

The electronic properties of graphene can be manipulated via mechanical deformations
, which opens prospects for both studying the Dirac fermions in new regimes and for new device applications. 
Certain natural configurations of strain generate large nearly uniform pseudo-magnetic fields, which have opposite signs in the two valleys, and give rise to flat spin- and valley-degenerate pseudo Landau levels (PLLs). Here we consider the effect of the Coulomb interactions in strained graphene with uniform pseudo-magnetic field. We show that the spin/valley degeneracies of the PLLs get lifted by the interactions, giving rise to topological insulator-like states. In particular, when a nonzero PLL is quarter- or three-quarter filled, an anomalous quantum Hall state spontaneously breaking time-reversal symmetry emerges. At half-filled PLL, weak spin-orbital interaction stabilizes time-reversal-symmetric quantum spin-Hall state. 
These many-body states are characterized by the quantized conductance and persist to a high temperature scale set by the Coulomb interactions, which we estimate to be a few hundreds Kelvin at moderate strain values. At fractional fillings, fractional quantum Hall states breaking valley symmetry emerge. These results suggest a new route to realizing robust topological insulator states in mesoscopic graphene.


\end{abstract}
\maketitle

{\bf Introduction.} Graphene provides a unique example of a 2D membrane hosting a high-mobility two-dimensional electron gas~\cite{CastroNeto09}. Mechanical deformations of the membrane strongly couple to the electron motion, generating local potentials~\cite{Kim08} and gauge fields~\cite{Iordanskii85,Sasaki05,Morpurgo06,Morozov06}. The interplay of the mechanical and electronic degrees of freedom have been mostly discussed in relation to the transport properties of graphene samples which are often locally strained and rippled~\cite{CastroNeto09}. More recently, it was realized that strain \emph{engineering} provides a promising way to control the band structure and the spectrum of graphene, as well as to create new devices~\cite{Pereira08,Ni09,Bao09}.

Non-uniform strains locally shift the positions of the $K,K'$ Dirac nodes in the opposite directions; this is equivalent to the local vector potential, and, in general creates local pseudo-magnetic field, which has opposite signs in the two valleys~\cite{Iordanskii85,Sasaki05,Morpurgo06,Morozov06}. This fact implies that the time-reversal symmetry (TRS) is preserved by strain, as well as that translational symmetry is broken, since the states in two valleys generally have different magnetic translation operators. Recently, it was shown that certain natural strain configurations lead to a large, nearly uniform pseudo-magnetic field~\cite{Guinea09-1,Guinea09-2}. This drastically modifies the spectrum of graphene, creating a sequence of quantized, fourfold (valley and spin) degenerate pseudo Landau levels (PLLs). Recently, PLLs were observed in locally strained graphene~\cite{Crommie10,Bocrath10};  the effective pseudomagnetic field was as high as 300 T~\cite{Crommie10}.

{\bf Main results.} We study the effects of the electron interactions on the physics of partially filled PLLs in strained graphene. We show that the approximate four-fold valley and spin degeneracy of the PLLs gets lifted, giving rise to incompressible topologically non-trivial states. We find that in $n\neq 0$ PLLs, where $n$ numbers the PLLs in the standard way~\cite{CastroNeto09}, three distinct types of states emerge -- anomalous quantum Hall (AQH) state~\cite{Haldane88} breaking TRS, TRS-preserving quantum spin-Hall state (QSH)~\cite{KaneMele05}, as well as a trivial insulating state.

The AQH and QSH states are characterized by chiral and counter-propagating edge states, respectively. The edge states are insensitive to the crystallographic edge type; this should be contrasted with the case of incompressible states corresponding to the complete filling of a PLL, which, in general, have gapless edge states only for particular edge orientations~\cite{Guinea09-1}.

The mechanism leading to the formation of the broken-symmetry states is reminiscent of the quantum Hall ferromagnetism (QHFM), for the case of graphene discussed in Ref.~\cite{Nomura06}. Similarly to QHFM, in our problem valley/spin split states are favored by the exchange interactions; however, owing to the different underlying symmetries, the properties of the resulting split states are very different from the usual QHFM states.
In the leading order the Coulomb Hamiltonian of a partially filled PLL has a $Z_2\times SU(2)$ symmetry; $Z_2$ is the time reversal, which interchanges valley and spin species, while $SU(2)$ corresponds to the spin rotation symmetry. This should be contrasted with the case of an $SU(4)$-symmetric QHFM, where any linear combination of the two valleys can be occupied~\cite{Nomura06}).

Below we focus on the case of non-zero PLL (the case $n=0$, in general, does not give rise to topologically non-trivial states, as discussed at the end of the paper; also, see Ref.~\cite{Ghaemi11}). At partial filling $f=1,3$ of the $n\neq 0$ PLL quartet, the ground state is an Ising-type valley ferromagnet.  The strong Ising-type anisotropy stems from the lack of the exchange interaction between the states in $K$ and $K'$ valleys, which results from the opposite chirality of Landau orbitals in $K$ and $K'$ valleys. The valley ferromagnet at $f=1,3$ breaks $Z_2$ TRS, and is characterized by the quantized Hall conductivity $\sigma_{xy}=\pm e^2/h$, with the $+(-)$ sign for the $K(K')$ valley polarization.

At partial filling $f=2$, a QSH state is stabilized by weak SO interactions.
The main effect of the SO interaction is to make the wave function orbitals of one pair of PLL related by TRS different from those of the other pair. Because of that, one of the pairs of PLLs has a lower exchange energy; the ground state at $f=2$ thus corresponds to filling that pair of PLL. Such a state can be qualitatively viewed as two filled LLs for spin-up and spin-down electrons, which experience opposite magnetic field; this state preserves TRS and is characterized by the quantized spin-Hall conductance and zero charge-Hall conductance. In order for the QSH state to be energetically favorable, the SO-induced exchange energy anisotropy should exceed weak anisotropy due to the lattice effects, which favors ferromagnetic spin-polarized state, and which is estimated  below.



The distinctive property of the interaction-induced QSH and AQH states is their robustness. They persist up to a temperature scale set by the Coulomb interaction, which, as we estimate below, can reach 200K at a moderate strain-induced pseudomagnetic field of $B=10\, {\rm T}$~\cite{Guinea09-1}. This is much larger than the typical temperature scales set by bare SO interaction, which characterize the QSH states in graphene~\cite{KaneMele05} and in HgTe-based quantum wells~\cite{Bernevig,Molenkamp}; for example, the latter persists only to temperatures of several Kelvin. We expect that the robustness of the interaction-driven QSH states will be advantageous for their manipulation and for the studies of their edge states properties. Potentially, QSH states in strained graphene could also be used in spintronics applications.

{\bf Model.} The general single-particle Hamiltonian of the strained graphene is given by:
\be\label{eq:single-particle}
H_{K(K')}=v_0 \psi^\dagger (\tilde p_x \sigma_x \tau_z+\tilde p_y \sigma_y)\psi,
  \,\, \tilde{p}_i=p_i \pm eA_i,
\ee
where $v_0$ is the Fermi velocity, and $K(K')$ correspond to the two valleys of the Dirac fermions, located in the opposite corners of the Brillouin zone; $\sigma_i, \tau_i$ are the Pauli matrices acting in the sublattice and valley spaces. The vector-potential is related to the strain gradient~\cite{Morpurgo06}. The above Hamiltonian does not contain spin-orbital terms, which will be included below.

We will be interested in the strain configurations that give rise to nearly uniform pseudo-magnetic field; such configurations were described in detail in Refs.~\cite{Guinea09-1,Guinea09-2}. The corresponding vector potential is given by
\be\label{eq:gauge_field}
A_{x}=-\frac{1}{2} B y, \,\, A_y=\frac{1}{2} B x .
\ee
Such pseudo-magnetic field breaks the spectrum of graphene into a sequence of PLLs with energies given by~\cite{CastroNeto09}:
\be\label{eq:LLs}
E_n={\rm sgn}(n) \sqrt{|n|} \epsilon_0, \,\,\, \epsilon_0=\frac{\sqrt{2}\hbar v_0}{\ell},
\ee
where $\ell=\sqrt{\frac{\hbar}{eB}}$ is the effective magnetic length set by $B$, and $v_0$ is is assumed to be constant, as appropriate for strains which are not too strong. In what follows, without loss of generality,  we take $B>0$ for definiteness.

In our analysis of the Coulomb interactions, we will need PLL wave functions. The details can be found in the Appendix; here we will just use the final result. The $n\neq 0$ PLL wave functions of the Hamiltonian in $K$ valley are given by:
\be\label{eq:LLwave_function}
\psi_{n,m}(z)=\frac{1}{\sqrt{2}}\left(\phi_{|n|-1,m}(z)  , \,\,  {\rm sgn}(n) \phi_{|n|,m}(z) \right),
\ee
where $\phi_{n,m}$ are the wave functions in the $n$th non-relativistic LL (see Appendix for the definition). 
The zeroth LL wave functions have a different form, with only bottom spinor component being non-zero:
\be\label{eq:LLwave_function0}
\psi_{n,m}(z)=\left( 0, \,\, \phi_{0,m}(z)
 \right).
\ee
The wave functions in the $K'$ valley have the same form as those in the $K$ valley (\ref{eq:LLwave_function}) (notice that this is different from the case of graphene in real magnetic field, where one would have to interchange the upper and lower spinor components). However, notice that the magnetic oscillator wave functions $\phi_{n,m}$ describe the opposite direction of the cyclotron motion compared to the $K$ valley. 

{\bf The Coulomb Hamiltonian.} In our analysis of the interaction effects, we will neglect the effects of Landau levels mixing. This amounts to projecting the density operators onto the partially filled $n$th LL. The effective Hamiltonian of the partially filled PLL can be written down in analogy with the case of real magnetic field (for a review, see Ref.~\cite{MacDonald94}):
\be\label{eq:coulomb}
H=\frac{1}{2S}\sum V({\bf q}) \rho({\bf q}) \rho(-{\bf q}), \,\, \rho({\bf q})=\sum_{\kappa s}\bar\rho_{\kappa s}({\bf q}),
\ee
where $\kappa=K,K'$, $s=\uparrow,\downarrow$ label valley and spin states, and $\bar\rho_{\kappa s}({\bf q})$ are the density operators projected onto the $n$th  LL. They have a different form in the $K$ and $K'$ valleys:
\be\label{eq:rho_projK}
\bar\rho_{K s} ({\bf q})=\sum_{m,m'} e^{-q^2/2}F_n({\bf q}) G_{m'm}(q) c_{Ks,nm'}^\dagger c_{Ks,nm},
\ee
\be\label{eq:rho_projK'}
\bar\rho_{K' s} ({\bf q})=\sum_{m, m'}e^{-q^2/2} F_n({\bf q}) G_{m'm} (\bar q) c^\dagger_{K' s,nm'}  c_{K' s, nm},
\ee
where we defined $q,\bar q=q_x\pm iq_y$, being measured in units of $1/\ell$, and $G$ is given by~\cite{MacDonald94} 
$$G_{m'm}(q)=(m!/m'!)^{1/2} (-iq/\sqrt 2)^{m'-m} L_m^{m'-m}(|q|^2/2),$$ $L_m^k$ being the generalized Laguerre polynomial; $F_n({\bf q})$ are the graphene form-factors, which encode the structure of PLL wave functions, and are identical to the case of magnetic field~\cite{Nomura06}
\be\label{eq:FF}
F_n({\bf q})=\frac{1}{2}\left[L_{|n|-1}(q^2/2)+L_{|n|}(q^2/2)\right].
\ee
The different structure of the projected density operators (\ref{eq:rho_projK},\ref{eq:rho_projK'}) reflects the fact that the pseudo-magnetic field has the opposite direction in the two valleys. This difference, although it may appear minor, has important implications for the energetics of the broken-symmetry states.

{\bf Competing phases.} Similarly to the case of QHFM, the physics of the partially filled PLL is governed by the exchange interactions. Below we show that the exchange energy is minimized when $f$ sub-levels in $K$ and/or $K'$ valley and with arbitrary spin projection are filled (that is, we are dealing with an Ising-type valley ferromagnet, and no linear combinations of the two valleys are allowed). This stems from the different cyclotron motion direction of the wave functions in $K,K'$, which causes the exchange contribution due to the correlations between different valleys to vanish~\cite{unpublished}. We also show that the ground state degeneracy is lifted by weak SO interactions, which break $SU(2)$ symmetry of the Coulomb Hamiltonian. This, generally, chooses a particular spin orientation at $f=1,3$, and favors QSH state at $f=2$.




 {\it Partial filling $f=1$.} We will start our analysis with the case of partial filling $f=1$. As we shall see, some of the results obtained here will be also useful for the case $f=2$.
 By analogy with the case of the quantum Hall ferromagnet, let us consider the following trial wave function:
 \be\label{eq:trial}
 |\Psi\ra=\prod_{m} d_m^\dagger |\Omega \ra,
 \ee
 where $|\Omega\ra$ is the vacuum, and $d$ is obtained by a unitary rotation $U$ of different valley and spin states:
 \be\label{eq:d}
 d^\dagger_m=\sum_{\kappa s} \bar U_{\kappa s} c^\dagger _{\kappa s, nm}.
 \ee

Evaluating the average of the Hamiltonian (\ref{eq:coulomb}) over the state (\ref{eq:trial}), and noting that the exchange integrals vanish between states in different valleys in a macroscopic sample, we find the exchange energy per particle in the relevant PLL~\cite{unpublished}:
\be\label{eq:energy1}
E=-\Delta_n \left[ (n_{K\uparrow}+n_{K\downarrow})^2+(n_{K'\uparrow}+n_{K'\downarrow})^2 \right],
\ee
where $n_{\kappa s}=U_{\kappa s}\bar U_{\kappa s}$ is the filling factor of the $\kappa s$ PLL, and
$$
\Delta_n=\frac{1}{2}\int \frac{d^2 q}{(2\pi)^2} V(q) F_n^2(q)e^{-q^2/2}.
$$
From the Eq.(\ref{eq:energy1}), we see that the lowest energy states are completely valley polarized, $n_{K\uparrow}+n_{K\downarrow}=1, n_{K'\uparrow}+n_{K'\downarrow}=0$ (or vice versa). 


The ground state is a valley ferromagnet that breaks TRS. Such a state is characterized by the quantized Hall conductivity and chiral edge states. The domain walls between domains with opposite valley polarization will also carry two chiral 1D modes.

What lifts the spin degeneracy of the ground state? The spin-orbital interactions, which entangle orbital and spin degrees of freedom, do not always lift the single-particle degeneracy of the Landau levels. However, they do modify the PLL wave functions for different spin projections (TRS is still preserved, such that there are two pairs of identical orbitals). This modifies the corresponding form-factors, breaking the $SU(2)$ invariance of the energy functional, and favoring some particular spin projection.

We illustrate this general effect by considering the intrinsic SO interaction~\cite{KaneMele05}:
\be\label{eq:SO}
H_{SO}=\Delta_{SO} \tau_z \sigma_z s_z,
\ee
where $s_i$ acts in the spin space. This type of SO interaction conserves $s_z$ component of spin, and therefore is the easiest to analyze; other types of SO have the same qualitative effect. 


The intrinsic SO (\ref{eq:SO}) modifies the wave functions in the $K$ valley (\ref{eq:LLwave_function}), making the spin-up and spin-down wave functions different:
\be\label{eq:LL_new}
\psi^{\uparrow}_{n,m}(z)=\left(
          \cos\xi \phi_{|n|-1,m}(z)  \,\,\,
          \sin\xi \phi_{|n|,m}(z)
 \right),
 \ee
 \be\label{eq:LL_new2}
\psi^{\downarrow}_{n,m}(z)=\left(
          \sin\xi \phi_{|n|-1,m}(z)  \,\,\,
          \cos\xi \phi_{|n|,m}(z)
 \right),
\ee
where $\tan\xi={\rm sgn}(n)\sqrt{1+(\Delta_{SO}/\epsilon_0)^2}-\Delta_{SO}/\epsilon_0.$
Similarly, the wave functions in the $K'$ valley will be modified, in such a way that the TRS is preserved: $K'\uparrow (K'\downarrow)$ state wave function will have the same form as the $K\downarrow (K\uparrow)$  wave function. Thus, there are now two pairs with different spatial wave functions.

The difference in the wave functions makes the form-factors of the two spin species different, leading to the difference in the exchange energy. The new form-factors are given by:
\be\label{eq:ff_new}
F_{n}^{K\uparrow}({\bf q})=F_{n}^{K'\downarrow}({\bf q})=\cos^2\xi L_{|n|-1}(q^2/2)+\sin^2\xi L_{|n|}(q^2/2),
\ee
\be\label{eq:ff_new2}
F_{n}^{K\downarrow}({\bf q})=F_{n}^{K'\uparrow}({\bf q})=\sin^2\xi L_{|n|-1}(q^2/2)+\cos^2\xi L_{|n|}(q^2/2).
\ee

This breaks the spin-rotational symmetry of the Coulomb Hamiltonian, generating an analogue of the Zeeman interaction in the exchange energy functional:
\be\label{eq:energy_anisotropic}
E=-\Delta_n (n_{K\uparrow}+n_{K\downarrow})^2-\delta_z (n_{K\uparrow}-n_{K\downarrow}),
\ee
where we assumed that the electrons have already been valley-polarized in $K$ direction by their strong Coulomb repulsion, and the Zeeman-like term is given by, to the leading order in $\Delta_{SO}/\epsilon_0$, 
\be\label{eq:pseudo-zeeman2}
\delta_z=\frac{\Delta_{SO}}{4E_n}\int \frac{d^2 q}{(2\pi)^2} V(q)e^{-\frac{q^2}{2}} \left( L_{|n|-1}^2 (q^2/2)-L_{|n|}^2 (q^2/2) \right).
\ee
For $|n|=1$, evaluating the integral, we obtain $\delta_z=\frac{\alpha\sqrt{\pi} \Delta_{SO}}{32}$, where $\alpha=e^2/\hbar v_0 \epsilon$ is the coupling constant. This anisotropy, although small, fixes the spin polarization to be $s_z=1$. Alternatively, a state with the same energy could be obtained by filling $K'\downarrow$ PLL. 

{\it Partial filling $f=2$.} Now we proceed to the case $f=2$. In the leading order (neglecting SO interaction), there is a manifold of degenerate ground states: any two non-equivalent PLLs ($\kappa s$, $\kappa' s'$, where $\kappa,\kappa'=K,K'$, and $s,s'$ are linear combinations of spin-up and spin-down species) can be filled. Similarly to the case $f=1$, this degeneracy is lifted by the SO interactions, which modify the PLL wave functions and reduce the symmetry of the Coulomb Hamiltonian. For the case of intrinsic SO, the exchange energy is minimized when a time-reversed pair of PLL, $K\uparrow, K'\downarrow$, is filled.

Such a state is TRS-preserving, and can be viewed as a combination of filled spin-up and spin-down LLs subject to an {\it opposite} magnetic field. Via a Laughlin-type argument, we conclude that such a state is characterized by two counter-propagating spin-filtered edge states~\cite{KaneMele05}, and a quantized spin-Hall conductivity, $\sigma_{SH}=\sigma^{\uparrow}_{xy}-\sigma^\downarrow_{xy}=2e^2/h$. Moreover, the gapless edge states are protected as long as TRS is preserved; this can be seen by employing Kane and Mele's $S$-matrix argument~\cite{KaneMele05}. This argument shows that the TRS forces the off-diagonal elements of the edge states $S$-matrix to be zero, and thus, backscattering is prohibited as long as TRS is not broken.

{\it Case of Rashba interaction.} Other types of SO interaction (for concreteness, we focus on the Rashba term), do not conserve $s_z$ projection. Despite this, their effect on the $f=2$ is similar to the $s_z$-conserving intrinsic SO: they break the spin rotational symmetry, making the wave functions of two time-reversed PLL pairs different~\cite{unpublished}. One of the pairs of PLLs has lower exchange energy, and the ground state corresponds to filling that pair.

Importantly, such a state is also a quantum spin-Hall insulator, with a pair of protected gapless edge states. This is because it can be adiabatically connected to the quantum spin-Hall insulator in the presence of only intrinsic SO interaction (when SO interaction is adiabatically changed from pure intrinsic type to the Rashba type, the exchange gap does not close). This should be contrasted with the Kane-Mele model~\cite{KaneMele05}, where the Rashba interaction weakens and eventually destroys the quantum spin-Hall state.

{\it $T_c$ estimate.} The critical temperature at which the interaction-induced topological states set it, is determined by the exchange gap, $T_c\sim \Delta_n$. For $n=1$ PLL we obtain $\Delta_n= \frac{11}{32}\sqrt{\frac{\pi}2} \frac{e^2}{\epsilon \ell}$, where we restored the dimensional factors. Taking $\epsilon\approx 5$ (the intrinsic screening of graphene~\cite{CastroNeto09})  gives $\Delta_n \approx 200 \, {\rm K}$ at $B=10 \, {\rm T}$. This scale far exceeds the bare SO interactions $\Delta_{SO}$; thus, we expect the interaction-induced QSH state discussed above to be significantly more robust than the QSH insulator in non-interacting systems~\cite{KaneMele05,Molenkamp}. Such a state will also show greater robustness with respect to disorder.

{\bf Concluding remarks}. Before we conclude, several comments are in order. First, much of the analysis above carries over to the splitting of the $n=0$ PLL, previously considered in Ref.~\cite{Ghaemi11}. At partial filling $f=1,3$ of the zeroth PLL, we expect valley-polarized states, while at $f=2$ weak SO interactions would favor TR-symmetric state (also, note that intrinsic SO splits $n=0$ PLL already on the single-particle level). However, in contrast to the case of $n\neq 0$ PLLs, we do not expect these states to exhibit protected edge states; this is because, in general, zeroth PLLs in $K (K')$ valleys cannot be characterized by a quantized Hall conductivity.

We also expect the electron interactions to give rise to fractional quantum Hall states, which break TRS spontaneously. In particular, we expect robust fractional states to emerge at partial fillings $f=k+1/3$, $f=k+2/3$, $k=0,1,2,3$, of the $n=0$ and $n=1$ PLLs~\cite{unpublished}, while higher PLLs form-factors favor charge-density-wave states. Although a possibility of TRS-preserving fractional states~\cite{Levin09} is intriguing, within the variational approach for the case of pure long-range Coulomb interaction we found no evidence that such states could be energetically favorable~\cite{unpublished}. Notice that TRS-preserving fractional states found in~\cite{Ghaemi11} required carefully chosen combination of lattice-scale interaction parameters. 

{\bf Discussion.} In summary, we have shown that nonzero pseudo Landau levels in strained graphene host a number of broken-symmetry states, including quantum Hall-like states which occur in the absence of an external magnetic field. Furthermore, weak SO interactions stabilize QSH states which preserve time-reversal symmetry.

We considered the effects of long-range Coulomb and SO interactions. The omitted short-range part of Coulomb interaction can stabilize the TRS-broken spin-polarized states at $f=2$. Taking the simplest Hubbard form of the short-range repulsion, $H_{U}=U\sum_{i=A,B}n_{i\uparrow}n_{i\downarrow}$, where the index $i$ labels graphene lattice sites, we obtain a mean-field estimate of the energy cost per particle (in the relevant pseudo LL) for QSH state. The result is $0.12U [{\rm eV}]$ meV, where we set $B=10$T. Since the energy gain due to SO interactions has to exceed this value,  for practical observation of the QSH state an artificial enhancement of the SO strength in graphene is likely to be necessary. This can be achieved, e.g., by graphene functionalization with adatoms~\cite{Alicea11,Ma}.

The proposed  states feature robust edge states (chiral in the case of AQH state, counter-propagating in the case of QSH state). We thus expect that the latter can be probed experimentally via the quantization of two-terminal conductance, much in the same way it was done for the QSH state in HgTe quantum wells~\cite{Molenkamp}. The interaction-induced gaps can also be probed in STM experiments.

{\bf Acknowledgements.} DAP was supported by Welch Foundation grant TBF1473 and by the ARO MURI on bioassembled nanoparticle arrays. 

\bibliography{paper}

\vspace{2cm}
 
\section{Appendix}

{\bf Landau level wave functions.} Here we write down PLL wave functions, which were necessary for the analysis of the interaction effects. The Hamiltonian of the $K$ valley in the presence of the gauge field (\ref{eq:gauge_field}) can be rewritten in the following form:
\be\label{eq:ham_magnetic}
H_K=\epsilon_0 \left[ \begin{array}{cc}
          0  & a \\
         a^\dagger & 0
     \end{array} \right],  
\ee
where we have introduced raising and lowering operators of the magnetic oscillator~\cite{MacDonald94}, which can be expressed via the complex coordinates $z,\bar z$:
\be\label{eq:aa}
a=\frac{-i}{\sqrt{2}} (2\ell \partial +\bar z/2\ell), \,\,  a^\dagger=\frac{i}{\sqrt{2}} (-2\ell \bar\partial + z/2\ell).
\ee
Following Ref.~\cite{MacDonald94}, we introduce ladder operators $b,b^\dagger$, which commute with $a,a^\dagger$ and therefore with the Hamiltonian:
\be\label{eq:bb}
b=\frac{1}{\sqrt{2}} (2\ell\bar\partial + z/2\ell), \,\,
b^\dagger=\frac{1}{\sqrt{2}} (-2\ell \partial + \bar z/2\ell).
\ee
The eigenstates of the (non-relativistic) magnetic oscillator are given by:
\be\label{eq:eigenstates}
|n,m\ra = \frac{\left( b^\dagger \right)^n \left( a^\dagger \right)^m }{\sqrt{m! \,\, n!}} |0\ra,
\ee
where the vacuum state, annihilated by both $a$ and $b$, is given $\phi_{0,0}(z)=\frac{1}{\sqrt{2\pi \ell^2}}e^{-z\bar z/4\ell^2}$. Let us also introduce a notation $\phi_{n,m}(z)=\la z | n,m\ra$ for the $|n,m\ra$ state wave function.

The $n\neq 0$ PLL wave functions of the Hamiltonian  (\ref{eq:ham_magnetic}) are given by:
\be\label{eq:LLwave_function}
\psi_{n,m}(z)=\frac{1}{\sqrt{2}}\left(\begin{array}{c}
          \phi_{|n|-1,m}(z)  \\
          {\rm sgn}(n) \phi_{|n|,m}(z)
     \end{array}
 \right).
\ee
The zeroth LL wave functions have a different form, with only bottom spinor component being non-zero:
\be\label{eq:LLwave_function0}
\psi_{n,m}(z)=\left(\begin{array}{c}
          0  \\
          \phi_{0,m}(z)
     \end{array}
 \right).
\ee

The spectrum in the valley $K'$ can be found in the same manner. One important difference is that the pseudomagnetic field has an opposite sign, and therefore the operators $a,b$ should be defined differently:
\be\label{eq:aaK'}
a=\frac{i}{\sqrt{2}} (2\ell \bar\partial + z/2\ell), \,\,  a^\dagger=\frac{i}{\sqrt{2}} (2\ell \partial - \bar z/2\ell),
\ee
\be\label{eq:bb}
b=\frac{1}{\sqrt{2}} (2\ell\partial + \bar z/2\ell), \,\,
b^\dagger=\frac{1}{\sqrt{2}} (-2\ell \bar\partial + z/2\ell).
\ee
 The wave functions, defined using new operators $a,b$ have the same form as those in the $K$ valley (\ref{eq:LLwave_function}) (notice that this is different from the case of graphene in real magnetic field, where one would have to interchange the upper and lower spinor components).

\end{document}